\documentclass[usenatbib]{article}
\usepackage{times,graphicx,natbib,a4wide}

\begin{document}

\title{Numerical Testing of The Rare Earth Hypothesis using Monte Carlo Realisation Techniques}
\author{Duncan H. Forgan$^1$ and Ken Rice$^1$}
\maketitle

\noindent $^1$ Scottish Universities Physics Alliance (SUPA) \\
Institute for Astronomy, University of Edinburgh \\
Royal Observatory Edinburgh \\
Blackford Hill \\
Edinburgh EH9 3HJ \\
UK \\\\
Tel: 0131 668 8359 \\
Fax: 0131 668 8416 \\
Email: dhf@roe.ac.uk \\

\begin{abstract}

\noindent The Search for Extraterrestrial Intelligence (SETI) has thus far failed to provide a convincing detection of intelligent life.  In the wake of this null signal, many ``contact pessimistic'' hypotheses have been formulated, the most famous of which is the Rare Earth Hypothesis.  It postulates that although terrestrial planets may be common, the exact environmental conditions that Earth enjoys are rare, perhaps unique. As a result, simple microbial life may be common, but complex metazoans (and hence intelligence) will be rare.  This paper uses Monte Carlo Realisation Techniques to investigate the Rare Earth Hypothesis, in particular the environmental criteria considered imperative to the existence of intelligence on Earth. \\

\noindent By comparing with a less restrictive, more optimistic hypothesis, the data indicates that if the Rare Earth hypothesis is correct, intelligent civilisation will indeed be relatively rare.  Studying the separations of pairs of civilisations shows that most intelligent civilisation pairs (ICPs) are unconnected: that is, they will not be able to exchange signals at lightspeed in the limited time that both are extant.  However, the few ICPs that are connected are strongly connected, being able to participate in numerous exchanges of signals.  This may provide encouragement for SETI researchers: although the Rare Earth Hypothesis is in general a contact-pessimistic hypothesis, it may be a "soft" or "exclusive" hypothesis, i.e. it may contain facets that are latently contact-optimistic.\\
 

\end{abstract} 

\newpage

\section{Introduction}

\noindent The attributes of the planet Earth are of critical importance to the existence and survival of life upon it.  In fact, it may be so finely tuned that few planets in the Galaxy share its life-friendly characteristics.  From this premise, it is almost inevitable to reach the conclusion that intelligent life (at least any that is predicated on evolution from complex metazoans) is also rare - perhaps unique to the planet Earth. \\

\noindent These ideas have been encapsulated in what is known as the Rare Earth Hypothesis \citep{rare_Earth}.  It can be summarised thus:

\begin{enumerate}
\item Simple life may be commonplace in the Universe.  The existence of extremophilic organisms in what were originally considered to be inhospitable regions (hydrothermal vents, acidic pools, toxic waste, deep in the Earth's crust) has shown the hardiness of simple life \citep{extremophiles,extremophiles_2}.  Indeed, these habitats are believed to be duplicated elsewhere in the Solar System - e.g. Mars \citep{methane_mars,methane_mars_2}, Europa \citep{ocean_europa}, Titan \citep{lake_titan}, Enceladus \citep{plume_enceladus,plume_enceladus_2} -  so it is still possible that "alien" life may be closer to home than once thought.
\item However, although simple life is resilient and adaptable, the evolution of complex animal life is extremely difficult.  For this to be achieved, there are certain criteria (hereafter referred to as \emph{the Earth Criteria}) that must be satisfied, in order for animals to thrive.
\end{enumerate}

\noindent A (non-exhaustive) list of the Earth Criteria follows:

\begin{itemize}
\item A planet within a critical range of orbital radii - the ``stellar habitable zone'' \citep{Hart_HZ,Kasting_et_al_93}
\item A star within a critical mass range (large enough to push the habitable zone outside the planet tidal locking radius, and small enough to provide sufficient energy while avoiding UV exposure)
\item A star located in a critical region of the Galaxy - the ``galactic habitable zone'' \citep{GHZ}
\item A planet within a critical mass range to maintain a suitable atmosphere
\item A planet with a stable low eccentricity orbit (to avoid extreme temperature changes).  This also requires a relatively large moon to provide axial stability \citep{anthropic_moon}.
\item A planet with sufficient raw materials to generate amino acids and proteins
\item A planet with suitable atmospheric composition, in particular the production of atmospheric oxygen - initially produced by cyanobacteria in Earth's early history \citep{earth_oxygen} 
\item A planet with plate tectonic activity to regulate atmospheric composition and the balance of carbon \citep{how_rare})
\item The presence of Jupiter to control the rate of comet and asteroid impacts - although this is now in question \citep{jupiter_foe_1,jupiter_foe_2,jupiter_foe_3}.
\end{itemize}

\noindent The weakness of this hypothesis rests in the (usually implicit) assumption that all the Earth Criteria are independent of each other.  Taking Jupiter as an example: asking whether Jupiter exists or otherwise in the Solar System is not meaningful, as planet formation is a complex, non-linear process: every planet in the Solar System owes its formation to its surounding environment, and therefore its planetary neighbours, through the dynamics of migration \citep{Earth_after_giants,Paardekooper_and_Papaloizou_08} planet-planet scattering \citep{Ford_and_Rasio_08,planet_planet}, resonances \citep{resonances}, and other secular phenomena (e.g. \citealt{Batgyin_and_Laughlin_08}).  Without Jupiter, the Earth as it is today may not have formed at all. \\  

\noindent This paper will investigate the influence of a subset of the Earth Criteria on the resulting distribution of inhabited planets, using Monte Carlo Realisation Techniques \citep{Vukotic_and_Cirkovic_07,Vukotic_and_Cirkovic_08,mcseti1}.  The paper is structured as follows: section \ref{sec:Method} will outline the methods used to simulate the distribution of life in the Galaxy; section \ref{sec:Inputs} will define the input parameters for the simulations run; the results are displayed in section \ref{sec:Results} and summarised in section \ref{sec:Conclusions}.

\section{Method}\label{sec:Method}

\noindent The numerical simulations are carried out using the Monte Carlo Realisation techniques outlined in \citet{mcseti1}, hereafter referred to as Paper I.  A brief summary of the method follows for completeness. \\

\noindent In essence, the method generates a Galaxy of \(N_*\) stars, each with their own stellar properties (mass, luminosity, location in the Galaxy, etc.) randomly selected from observed statistical distributions.  Planetary systems are then generated for these stars in a similar manner, and life is allowed to evolve in these planets according to some hypothesis of origin.  The end result is a mock Galaxy which is statistically representative of the Milky Way.  To quantify random sampling errors, this process is repeated many times: this allows an estimation of the sample mean and sample standard deviation of the output variables obtained.  \\

\noindent The inputs used to define the mock Galaxy (e.g. the Galaxy's surface density profile, the initial stellar mass function (IMF), the star formation history (SFH), etc.) are of critical importance.  Paper I focussed on using current empirical data (especially for the simulation of exoplanets) to define the mock Galaxy.  This paper will attempt to improve on the inputs of Paper I: these improvements are discussed below.

\subsection{Improvements on the Model}

\subsubsection{The Simulation of the Galaxy}

\noindent In Paper I, the Milky Way was simulated in two dimensions only (in polar coordinates \((r,\phi)\)).  As a first improvement, the Galaxy is given vertical structure, incorporating both the thick and thin stellar discs \citep{Ostlie_and_Caroll}:

\begin{equation} \rho (r,z) = n_0 e^{-r_{gal}/r_{H}} \left(e^{-z_{gal}/z_{thin}} + 0.02e^{-z_{gal}/z_{thick}}\right)\end{equation}

\noindent Secondly, the metallicity gradient of the Milky Way was previously simulated using only one curve:

\begin{equation} Z_* = -z_{grad}\log\left(\frac{r_{gal}}{r_{gal,\odot}}\right) \end{equation}

\noindent In truth, there are many differing measurements of the abundance gradient in the Galaxy \citep{z_grad}, dependent on the metals studied.  This reflects the different synthesis processes at work for differing elements.  This can be (crudely) reproduced by allowing \(z_{grad}\) to have a distribution of values - in this case, a Gaussian distribution, with sample mean and sample standard deviation defined by the measurements of \citet{z_grad}. \\

\begin{figure}
\begin{center}
\includegraphics[scale = 0.5]{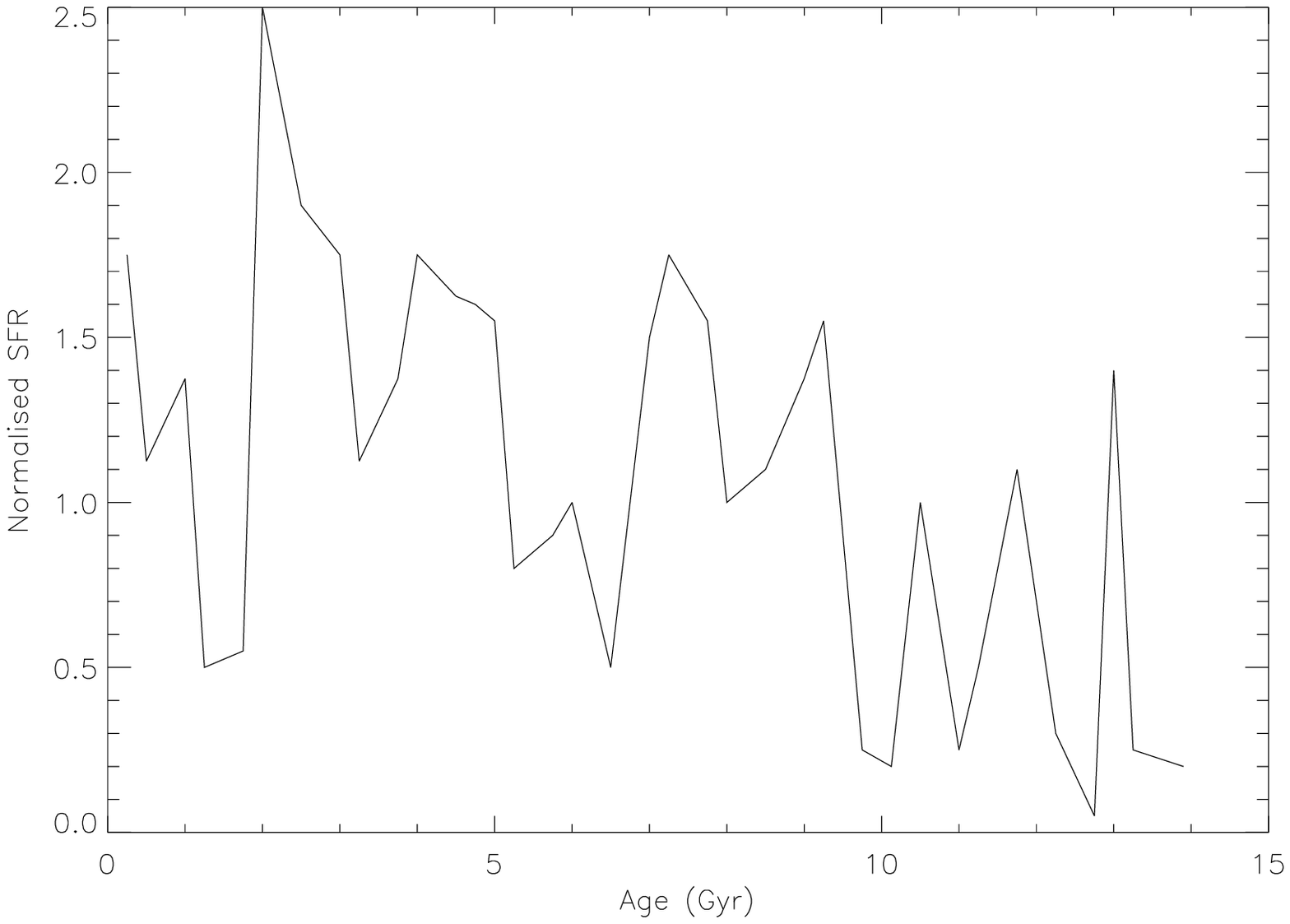}
\caption{\emph{The star formation history used in this work \citep{Rocha_Pinto_SFH}} \label{fig:SFH}}
\end{center}
\end{figure}

\noindent Finally, measures have been taken to correlate the age and metallicity of the stars.  The Age Metallicity Relation (AMR) of \citet{Rocha_Pinto_AMR} (with its errors) defines upper and lower bounds to the  age of a star (given its metallicity).  The star formation history (SFH) has also been improved, allowing better time resolution (\citealt{Rocha_Pinto_SFH}, see Figure \ref{fig:SFH}).

\subsubsection{The Simulation of Stars}

\noindent An important change to this work is the simulation of stellar luminosity evolution.  As stars evolve along the main sequence, their luminosity increases \citep{Schroder_and_Smith_08}.  As the luminosity increases, the location of the stellar habitable zone must move further away from the star \citep{Hart_HZ,Kasting_et_al_93}.  This implies that any planets with biospheres initially in the habitable zone can leave the habitable zone on a timescale \(\tau_{HZ}\), which is a function of the host star's initial luminosity and the planet's distance from it.  Together with the main sequence lifetime \(\tau_{MS}\), it defines a maximum lifetime for any biosphere:

\begin{equation} \tau_{max} = MIN(\tau_{MS}, \tau_{HZ}) \end{equation}

\noindent Where \(\tau_{MS}\) is given by

\begin{equation}\frac{t_{MS}}{t_{MS,\odot}} =\left[\frac{M_*}{M_{\odot}}\right]^{-3}  \end{equation}

\noindent For more information, see \citet{Prialnik} (Note that quantities referring to the Sun have the subscript \(\odot\)).  The luminosity evolution of the stars are approximated by extrapolating the simulated solar luminosity data of \citet{Schroder_and_Smith_08} to all main sequence stars\footnote{This may seem a weak assumption, but the stars of interest in these simulations will be close to solar type, so the approximation is reasonable in this first instance.}:

\begin{equation} L(t) = \left(0.7 +0.144\left(\frac{t}{Gyr}\right)\right)\frac{L_*}{L_{\odot}} \end{equation}

\subsubsection{The Simulation of Planets}

\noindent Current exoplanet data, while impressive, is still incomplete.  This introduces significant bias into the results of any simulation \citep{mcseti1}, and precludes the discussion of the Rare Earth Hypothesis if using observations alone (as statistical analyses cannot currently simulate Earths with any robustness).  To bypass this problem, the empirical data is replaced by theoretical relations: this allows the simulation of planetary objects down to lunar masses. \\

\noindent The probability of a star hosting planets is a function of its metallicity: this code uses the distribution as described by \citet{Wyatt_Z}:

\begin{equation} P(z) = 0.03 \times 10^{\frac{Z}{Z_{\odot}}} \end{equation}

\begin{figure}
\begin{center}
\includegraphics[scale = 0.5]{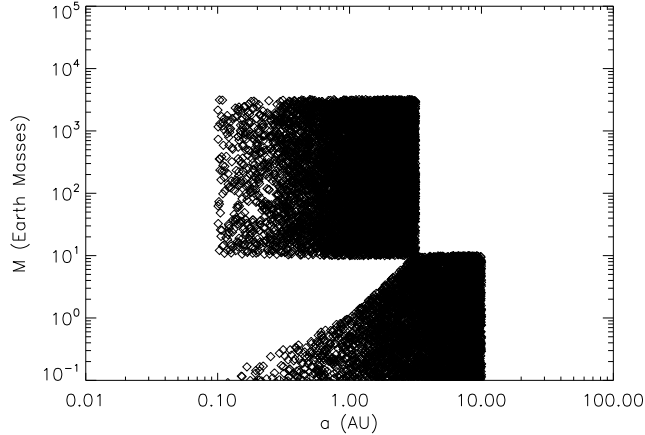}
\caption{\emph{The mass-radius relation for a sample of planets.} \label{fig:MvsR}}
\end{center}
\end{figure}

\noindent The Planetary Initial Mass Function (PIMF) is approximated by a simple power law:

\begin{equation} P(M_P) = (M_P)^{-1}\end{equation}

\noindent which operates over the mass range of \(\left[M_{moon}, 25\,M_{Jup}\right]\).  To correctly reproduce the distribution of planetary radii, two different radii distribution functions are used.  Jovian planets reproduce the data of \citet{Armitage_mig}, which accounts for the effects of Type II planetary migration.  For terrestrial planets, the data of \citet{Ida_Lin_mig} (Fig. 1c) is emulated: a simple parametrisation allows the trend for low mass objects to be recovered (see Figure \ref{fig:MvsR}).  It should be noted that in essence this is swapping one weakness for another: while the bias of empirical data is lost, the uncertainty of current planet formation models is gained. \\

\noindent Also, as the mass function can simulate Moon-mass objects, any object with a mass less than Pluto's that resides within another planet's Hill sphere (that is, it resides within the gravitational influence of said planet) is considered a moon of that planet\footnote{This does not guarantee the observational rule of thumb that the probability of a terrestrial planet having a substantial moon is around 0.25.  However, this is a reasonable first approximation, with more detailed studies of this issue requiring future observational input (e.g. \citealt{moon_detect,Kipping_moon}).}.  This property will become important for the simulation of the Rare Earth Hypothesis.

\subsubsection{The Simulation of Life}

\noindent The simulation of life proceeds in much the same manner as Paper I: life must achieve several difficult goals, each goal requiring a time \(\tau_i\) to be achieved, in order to evolve into a technologically capable, sentient species \citep{stages}.  During this process, resetting events can occur (e.g. cometary impacts, supernovae, gamma ray bursts, see \citealt{Annis}), which cause large scale extinctions, and reset the evolution of life to an earlier stage (or in the worst case, sterilise the planet entirely).  The reader is referred to Paper I for a more in depth discussion of how this is achieved in the code.  A significant change to the model is the mechanism for resetting events becoming sterilisation events.  This was described by a probability of annihilation \(p_{annihilate}\), which was created in order to define the Galactic Habitable Zone \citep{GHZ}.  This simplified approach has been replaced with an attempt to simulate the loss of biodiversity incurred as the result of a reset.  The effect of a reset is to reduce biodiversity by a fraction \(x\): if \(x\) is greater than 1, the planet is sterilised.  Appealing to the Central Limit Theorem, \(x\) is selected from a Gaussian Distribution with mean 0.5, and standard deviation 0.25.  This means that on average, 5\% of resets will result in annihilation.  The average number of resets increases with proximity to the Galactic Centre - this is now parametrised by:

\begin{equation} \mu_{resets} = \mu_{Earth} e^{-\left(\frac{r_{gal}}{r_{gal,\odot}}\right)} \end{equation}

\noindent Where \(\mu_{Earth} = 5\) \citep{Raup_and_Sepkoski}.  This provides the mean of a Gaussian distribution from which \(N_{resets}\) is sampled.  The \emph{habitation index} of Paper I (which is assigned to all planets in the simulation) is modified to account for the potential existence of microbial life:

\begin{equation} I_{inhabit} = \left\{
\begin{array}{l l }
-1 & \quad \mbox{Biosphere which has been annihilated} \\
0 & \quad \mbox{Planet is lifeless} \\
0.5 & \quad \mbox{Planet has microbial life} \\
1 & \quad \mbox{Planet has primitive animal life} \\
2 & \quad \mbox{Planet has intelligent life} \\
3 & \quad \mbox{Planet had intelligent life, but it destroyed itself} \\
4 & \quad \mbox{Planet has an advanced civilisation} \\
\end{array} \right. \end{equation}

\noindent Note that interplanetary colonisation is not modelled in this work.

\subsection{New Outputs - Civilisation Interaction}

\noindent As the code produces data pertaining to each individual civilisation, it is possible to study the entire dataset for each run, and identify the potential for communication between all possible pairs of civilisations.  For $N$ galactic civilisations, there are $N(N-1)/2$ pairs of civilisations.  For each intelligent civilisation pair (ICP), the following outputs can be calculated:

\begin{enumerate}
\item Their physical separation in kiloparsecs ($dx$),
\item The available window of communication $dt$ (that is, the maximum time interval where both civilisations exist and are able to communicate),
\item The space time interval $ds^2 = c^2dt^2 - dx^2$.  This quantity determines whether a signal travelling at lightspeed can traverse the distance between two civilisations within the communication window (assuming the intervening space to be Minkowskian).  If $ds^2 < 0$, then the signal will fail to reach its destination before the window closes.  If $ds^2=0$, then the signal will reach its destination at the same instant the window closes.  If $ds^2>0$, then the signal will reach its destination within the window, and it is therefore possible for communication between the two civilisations to be established.
\item The ``contact factor'' $f_{contact} = \frac{2c dt}{dx}$, which counts how many ``conversations'' (pairs of signals) can travel between the two civilisations.
\end{enumerate}  

\noindent These outputs give extra information on the distribution of civilisations in the Galaxy, and their potential connectedness by signals travelling at lightspeed.  

\section{Inputs}\label{sec:Inputs}

\noindent Two separate hypotheses were tested with this model.  Each was subjected to 30 Monte Carlo Realisations (MCRs), with each realisation containing $N_{stars} = 10^9$.  This is of course two orders of magnitude short of the Milky Way's stellar content, but computational constraints prevented increasing $N_{stars}$ further.  The interested will be able to multiply subsequent results by 100 to obtain an estimate of Milky Way figures.  In any case, absolute numbers are less relevant to the issue at hand: this study focuses on comparing two hypotheses, and comparing \emph{relative trends} (which is a more reliable route in studies of this nature).

\subsection{The Baseline Hypothesis}
 
\noindent This basic hypothesis requires only that a planet must be in the stellar habitable zone for life to form upon it.   If the planet's surface temperature lies between $[0,100]^{\circ}C$, then microbial life can form upon it.  Complex animal life will only form if the planet's surface temperature lies between $[4,50]^{\circ}C$ \citep{rare_Earth}.  This hypothesis was tested to provide a comparison with the results of the Rare Earth Hypothesis.

\subsection{The Rare Earth Hypothesis \label{sec:rare}} 

\noindent This hypothesis builds on the baseline hypothesis by also requiring that \emph{animal} life will only form on a planet if the following four conditions are met:

\begin{enumerate}
\item The planet's mass is between $[0.5,2.0]M_{\oplus}$, 
\item The star's mass is between $[0.5,1.5]M_{\odot}$,
\item The planet has at least one moon, (for axial stability and tides)
\item The star system has at least one planet with mass $>10M_{\oplus}$ in an outer orbit (for shepherding asteroids)
\end{enumerate} 

\section{Results \& Discussion}\label{sec:Results}

\subsection{The Distributions of Life and Intelligence}

\begin{figure*}
\begin{center}$
\begin{array}{cc}
\includegraphics[scale = 0.5]{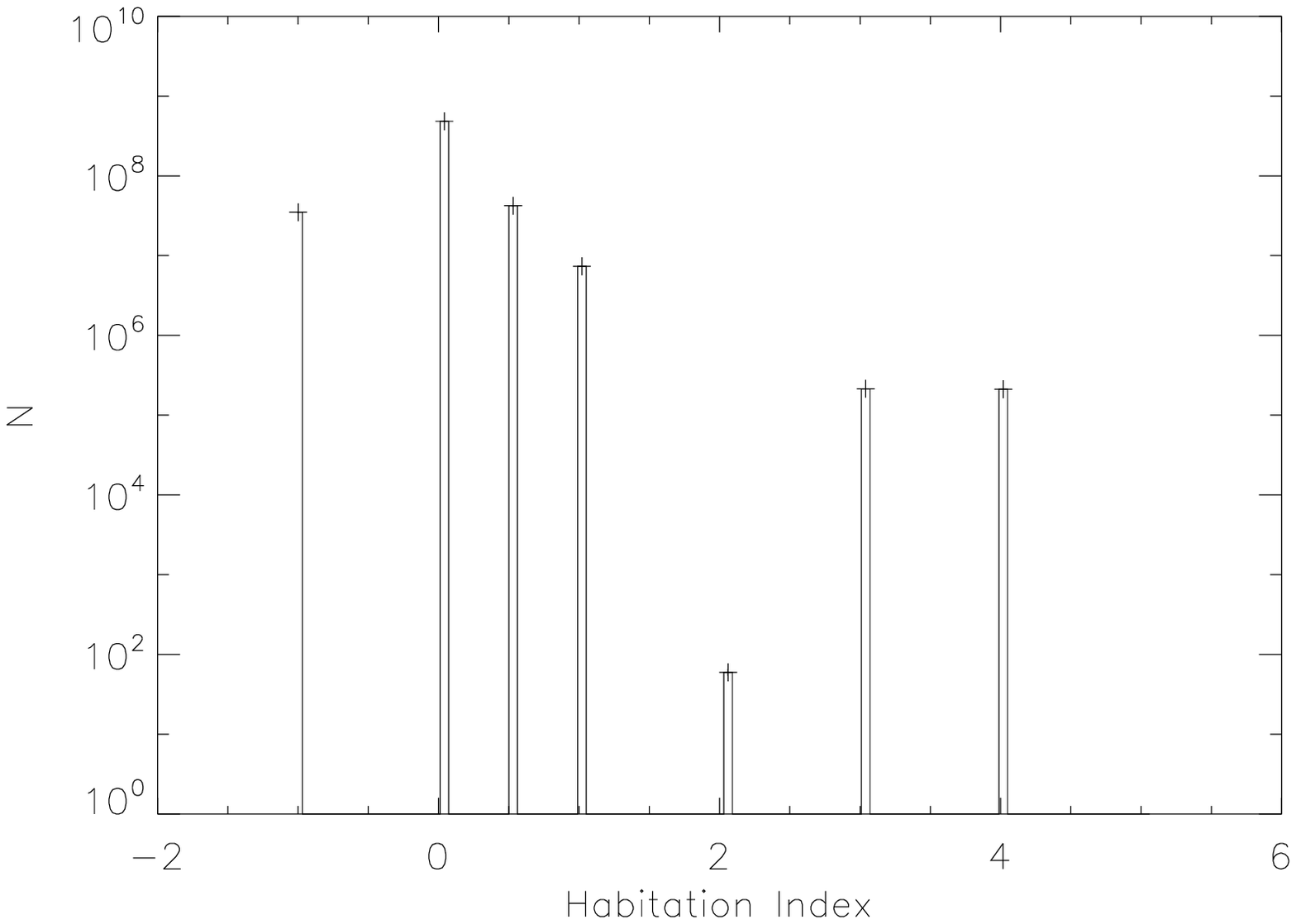} &
\includegraphics[scale = 0.5]{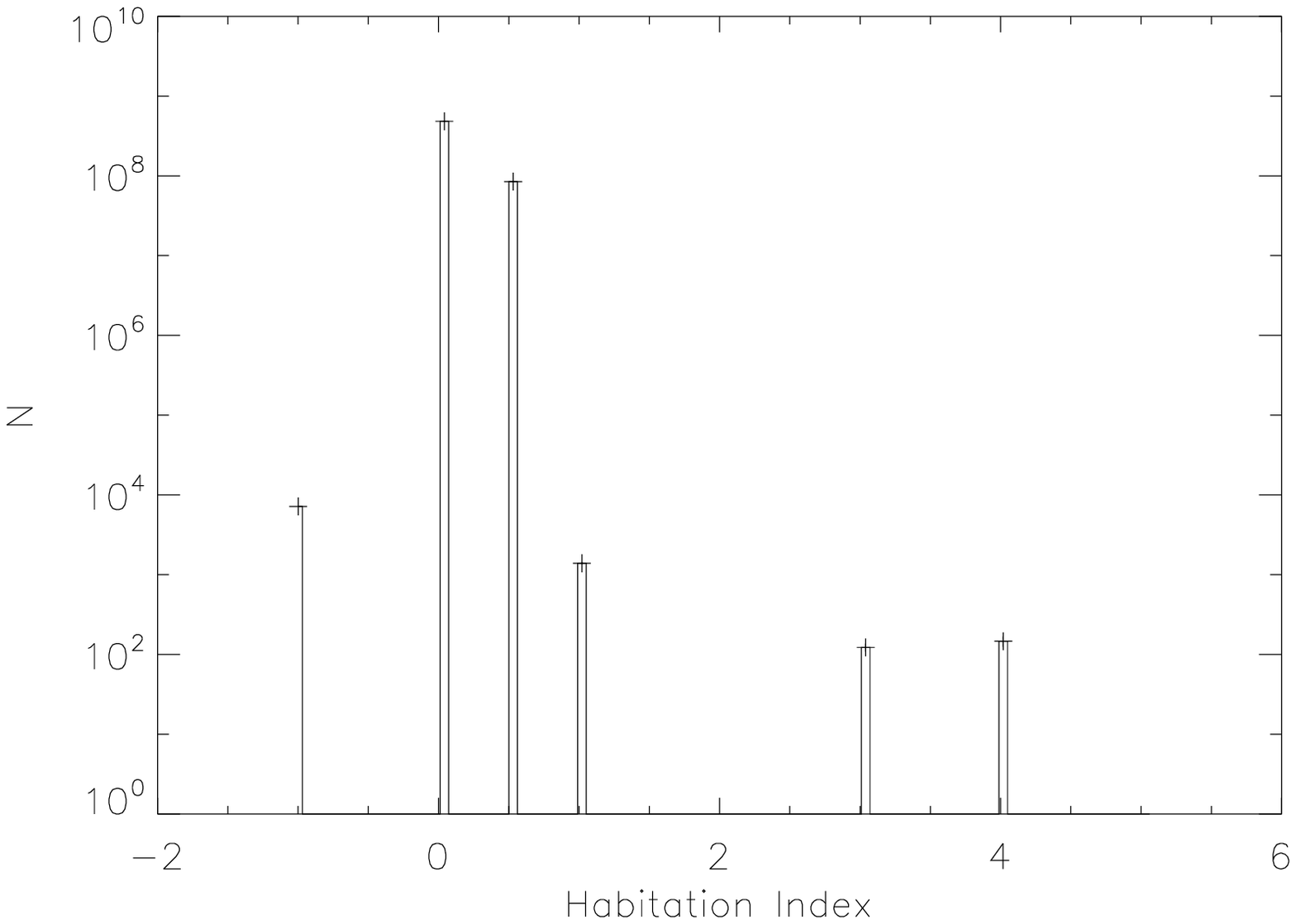} \\
\end{array}$
\caption{\emph{The habitation index for the Baseline Hypothesis (left) and the Rare Earth Hypothesis (right).} \label{fig:inhabit}}
\end{center}
\end{figure*}

\noindent The properties of the Rare Earth hypothesis galaxy can now be compared against those of the Baseline hypothesis.  Comparing the habitation index for both hypotheses (Figure \ref{fig:inhabit}), it can be seen that (by construction) microbial life ($I_{inhabit}=0.5$) is unaffected by the Rare Earth hypothesis, whereas the prevalence of animal life ($I_{inhabit} = 1$) is reduced by a factor of $10^4$ against the baseline.  This reduction is thus propagated into the intelligent biospheres ($I_{inhabit}>2$).  However, despite some quite stringent conditions on the planetary system architecture (conditions 3 and 4, section \ref{sec:rare}) the number of intelligent biospheres numbers in the hundreds: the implications for SETI are discussed in the next section. \\

\begin{figure*}
\begin{center}$
\begin{array}{cc}
\includegraphics[scale = 0.5]{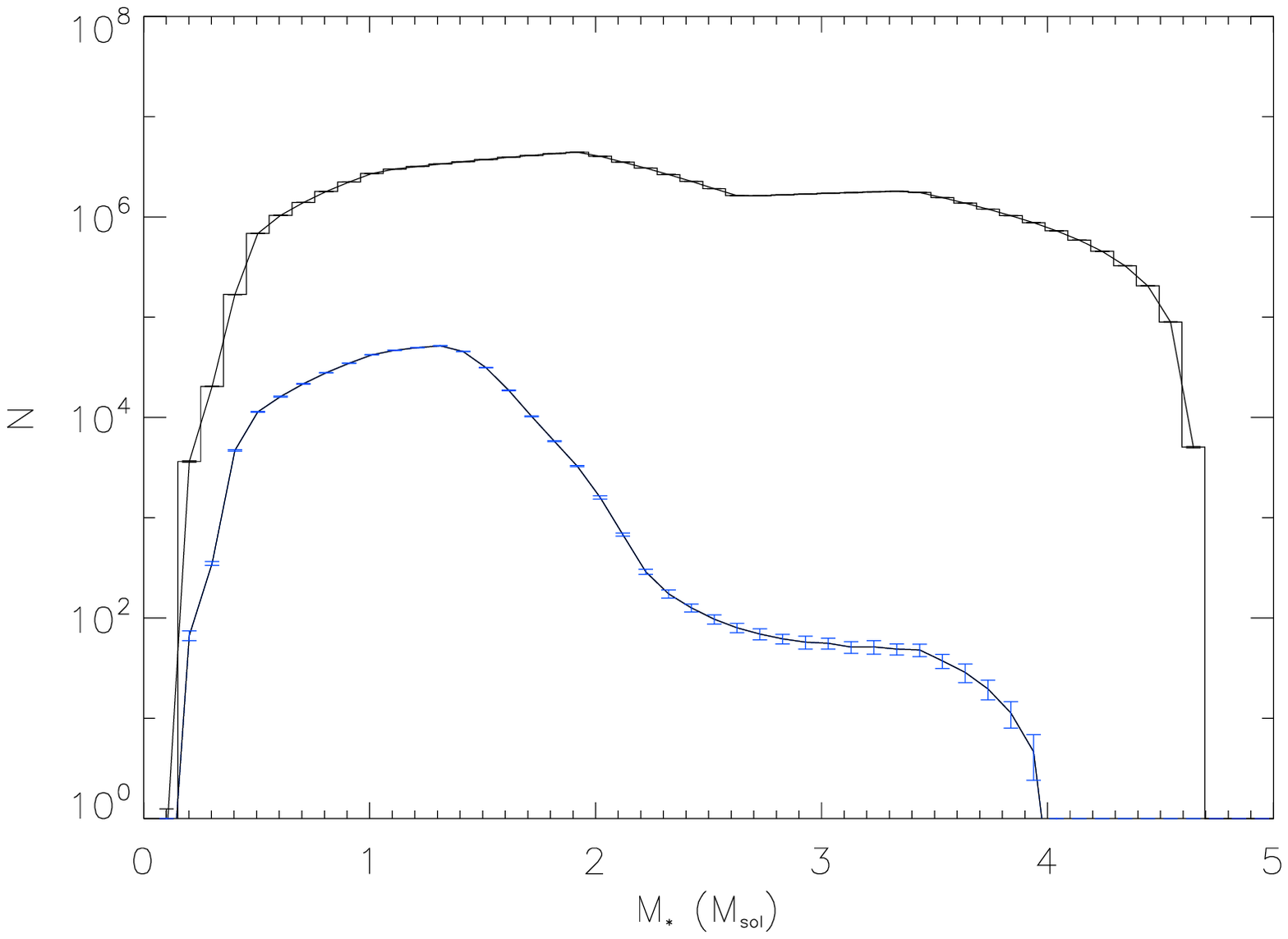} &
\includegraphics[scale = 0.5]{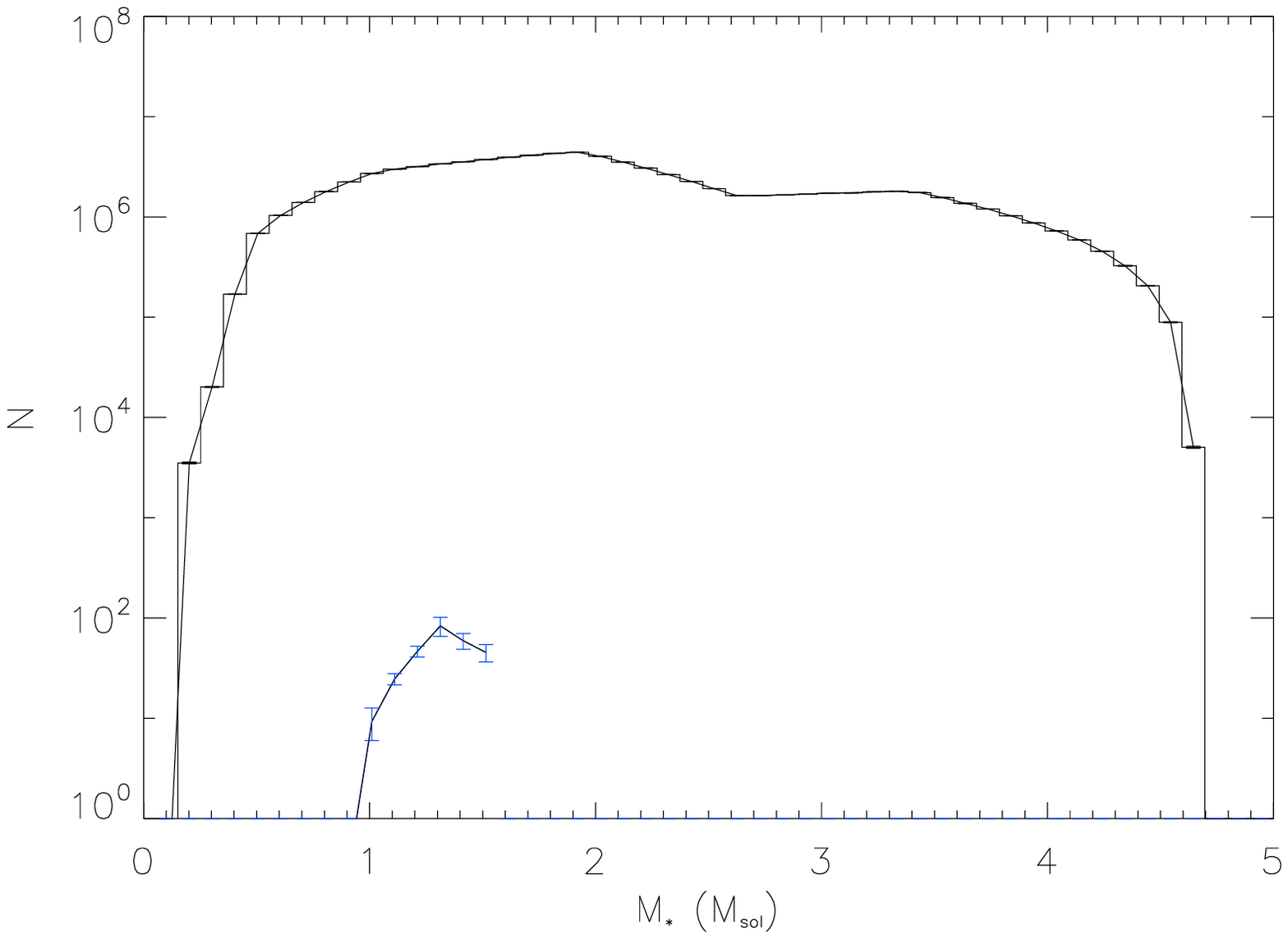} \\
\end{array}$
\caption{\emph{Stellar Mass for the Baseline Hypothesis (left) and the Rare Earth Hypothesis (right).  The black lines indicate all biospheres, the blue lines indicate all intelligent biospheres.} \label{fig:mstar}}
\end{center}
\end{figure*}

\noindent As stellar mass is a key condition to the Rare Earth Hypothesis, it should be expected that the two hypotheses' distributions diverge, and this is indeed the case: Figure \ref{fig:mstar} shows the distribution of stellar mass for both hypotheses.  The IMF is modified by the effects of the habitable zone (and the distribution of exoplanet semi-major axis) to give the characteristic bump between 1 and 2 solar masses. Comparing the hypotheses shows that although the Baseline hypothesis favours lower mass stars for intelligent biospheres (for their increased longevity), the Rare Earth Hypothesis must discard the substantial number of stars that are less than $0.5M_{\odot}$.  \\

\begin{figure*}
\begin{center}$
\begin{array}{cc}
\includegraphics[scale = 0.5]{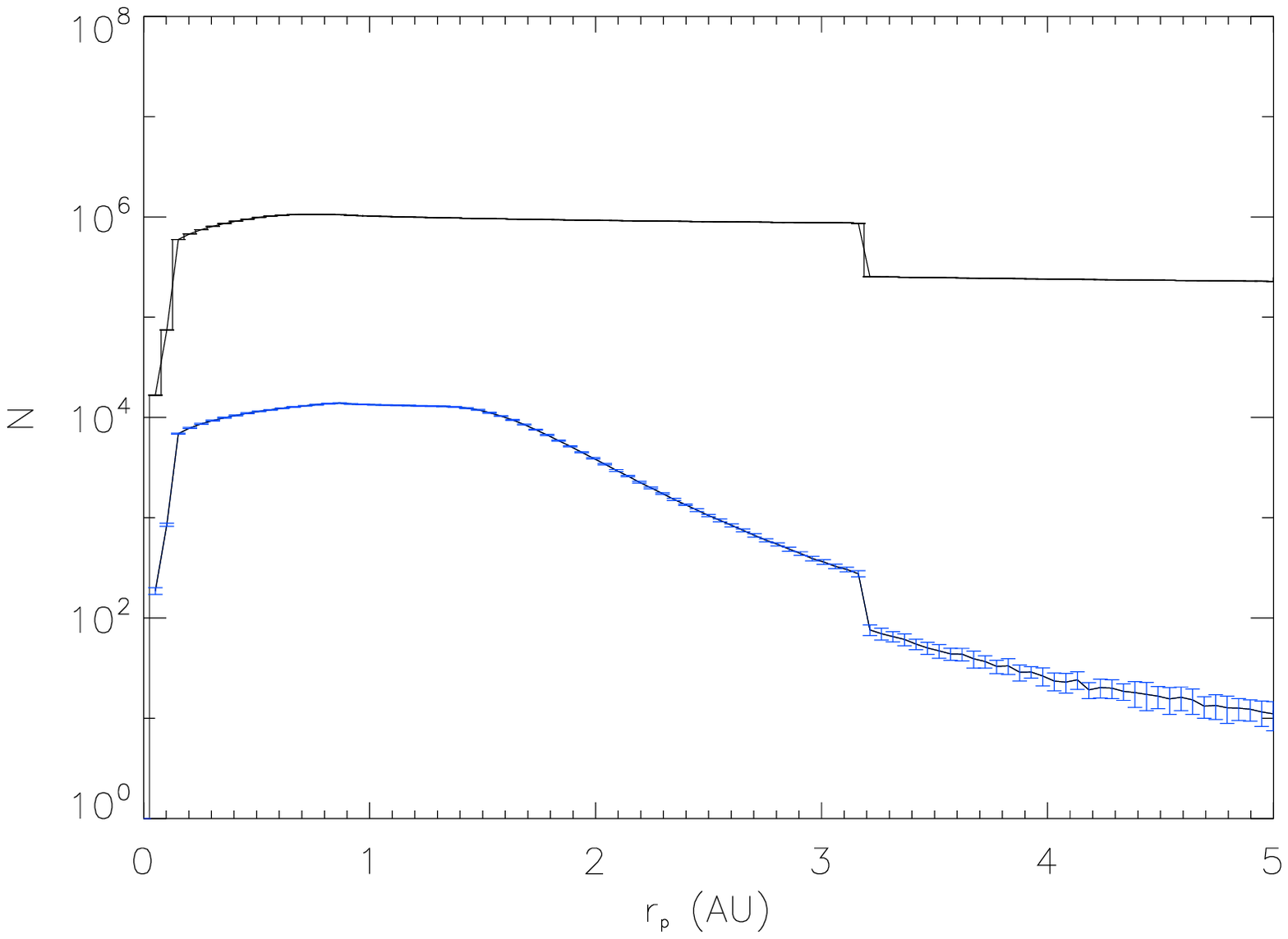} &
\includegraphics[scale = 0.5]{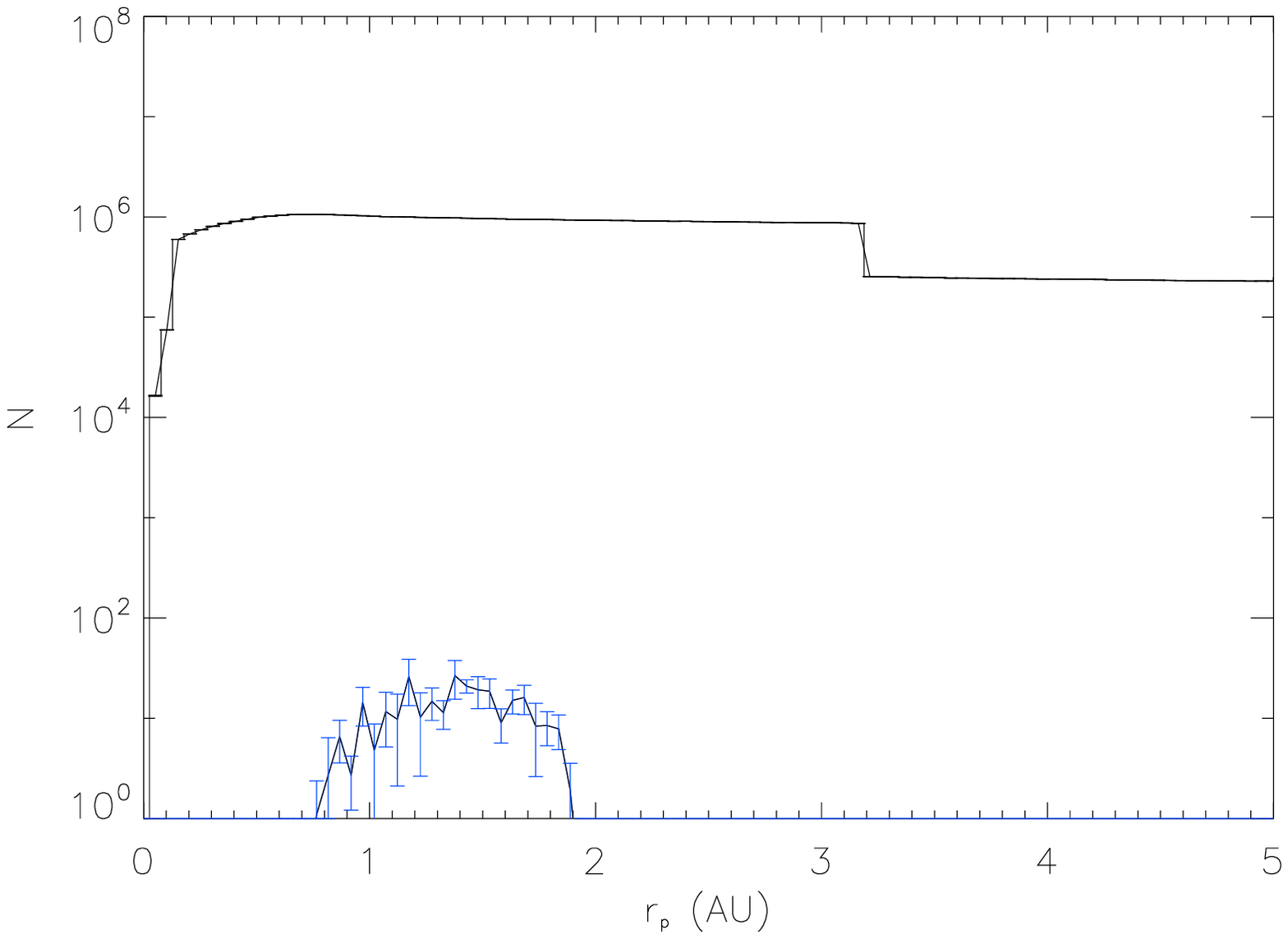} \\
\end{array}$
\caption{\emph{Planet semimajor axis for the Baseline Hypothesis (left) and the Rare Earth Hypothesis (right).  The black lines indicate all biospheres, the blue lines indicate all intelligent biospheres.} \label{fig:rp}}
\end{center}
\end{figure*}

\noindent This bias towards lower mass should be reflected in the distribution in semimajor axis (as lower mass stars have closer, more stationary habitable zones).  Figure \ref{fig:rp} shows that this is true for the Baseline Hypothesis (with intelligent biospheres dropping off as $R>1.5 \,AU$), and doubly true for the Rare Earth Hypothesis, selecting a narrow radial range between $[0.8,1.9] \,AU$.  \\

\begin{figure*}
\begin{center}$
\begin{array}{cc}
\includegraphics[scale = 0.5]{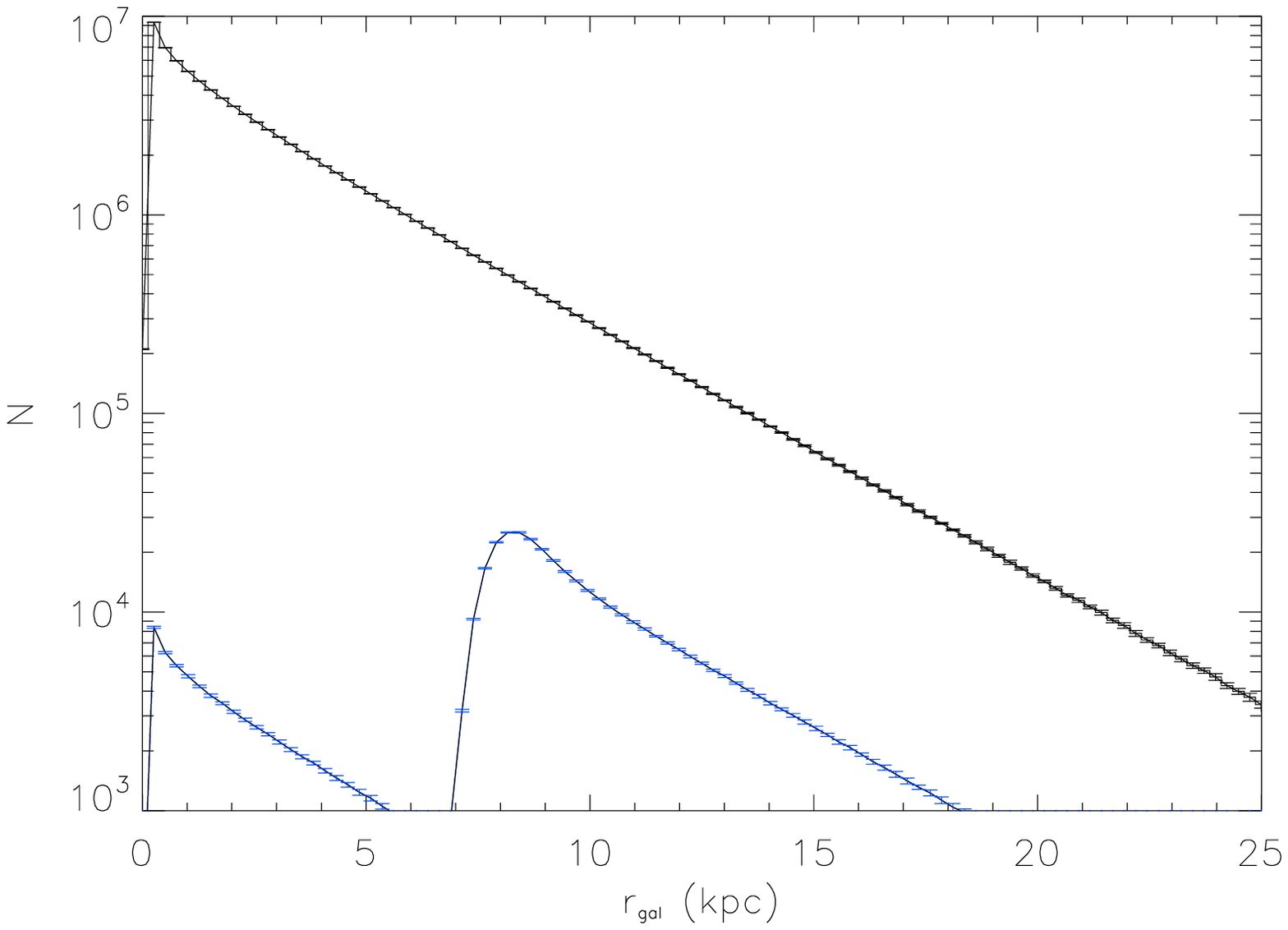} &
\includegraphics[scale = 0.5]{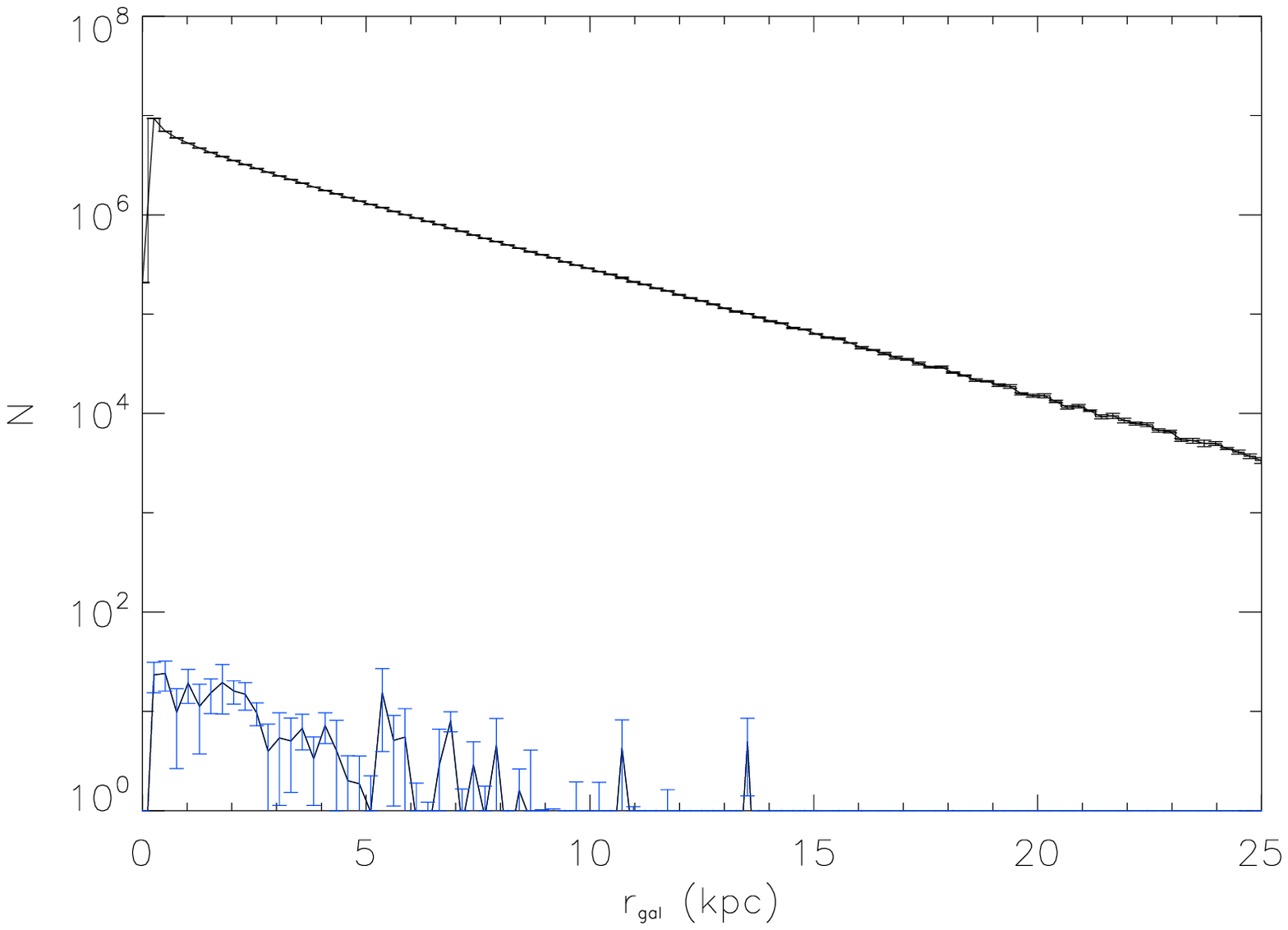} \\
\end{array}$
\caption{\emph{Galactocentric radius for the Baseline Hypothesis (left) and the Rare Earth Hypothesis (right).  The black lines indicate all biospheres, the blue lines indicate all intelligent biospheres.} \label{fig:rgal}}
\end{center}
\end{figure*}

\noindent The most striking difference can be seen in the distribution of galactocentric radius (Figure \ref{fig:rgal}).  While the Galactic Habitable Zone (GHZ) can be identified in the Baseline Hypothesis (with a small contingent at lower radii, which presumably exists due to the lack of modelling of the central supermassive black hole (SMBH) and hypervelocity stars in the inner regions), the Rare Earth Hypothesis appears to have no GHZ.  This is unexpected: the four conditions of the Rare Earth Hypothesis tested here do not affect where intelligent systems should lie; why then does the GHZ not appear (with reduced numbers)? \\

\subsection{Communication and Connectivity}

\begin{figure*}
\begin{center}$
\begin{array}{cc}
\includegraphics[scale = 0.5]{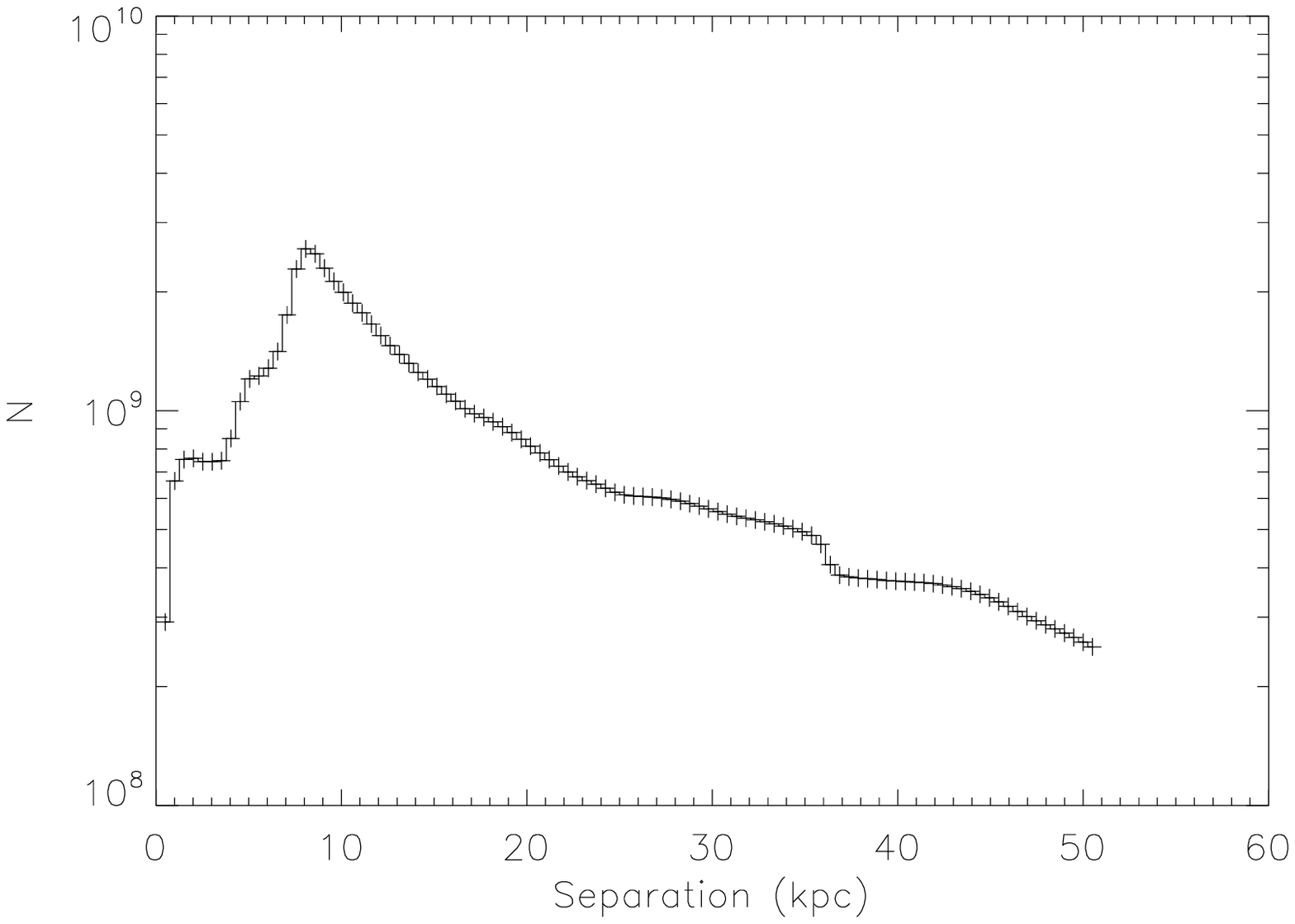} &
\includegraphics[scale = 0.5]{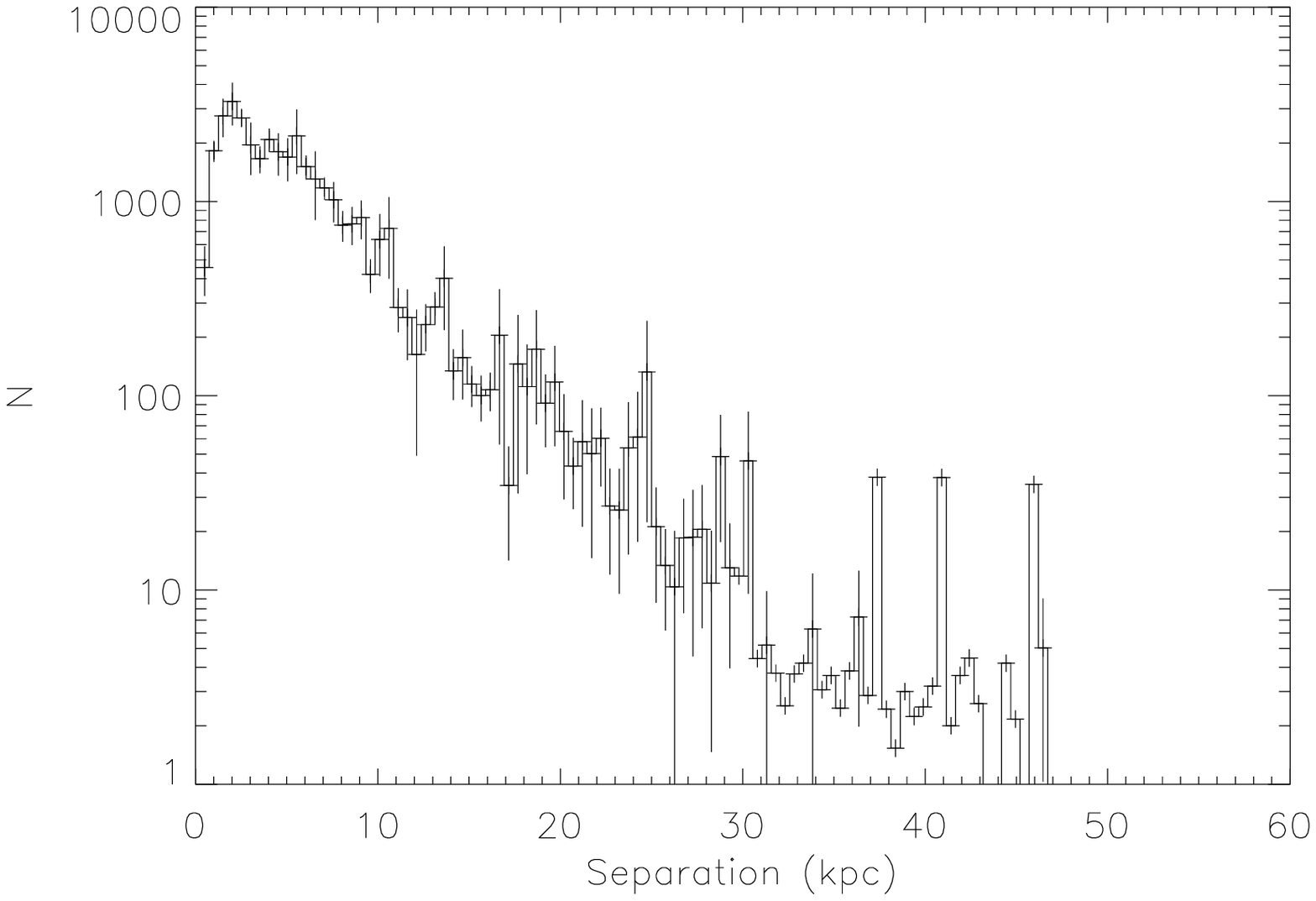} \\
\end{array}$
\caption{\emph{Separations of ICPs for the Baseline Hypothesis (left) and the Rare Earth Hypothesis (right).} \label{fig:dx}}
\end{center}
\end{figure*}

\noindent The \emph{prima facie} conclusion (having studied the results of the previous section), is that if the Rare Earth Hypothesis is correct, and intelligent civilisations are infrequent, then the potential for communication is also low.  This expectation can be tested by calculating the interaction variables discussed previously.  As the focus has now shifted from individual intelligent civilisations to intelligent civilisation pairs (ICPs), the numbers duly increase from $N$ to $\frac{N(N-1)}{2}$.  Figure \ref{fig:dx} shows the distribution of ICP separation $dx$ for both hypotheses.  The baseline hypothesis exhibits a sharp peak at around 8 kpc (the location of the GHZ), accompanied by a long decay.  This distribution is reminiscent of the lognormal distribution expected if the tools of the Statistical Drake Equation were applied \citep{stat_drake}.  The distribution reaches its mode in steps: these steps are presumably sensitive to the local galactic spiral structure.  The Rare Earth Hypothesis has no apparent GHZ, so the distribution (though reduced in magnitude) peaks at a much lower 3 kpc. \\

\begin{figure*}
\begin{center}$
\begin{array}{cc}
\includegraphics[scale = 0.5]{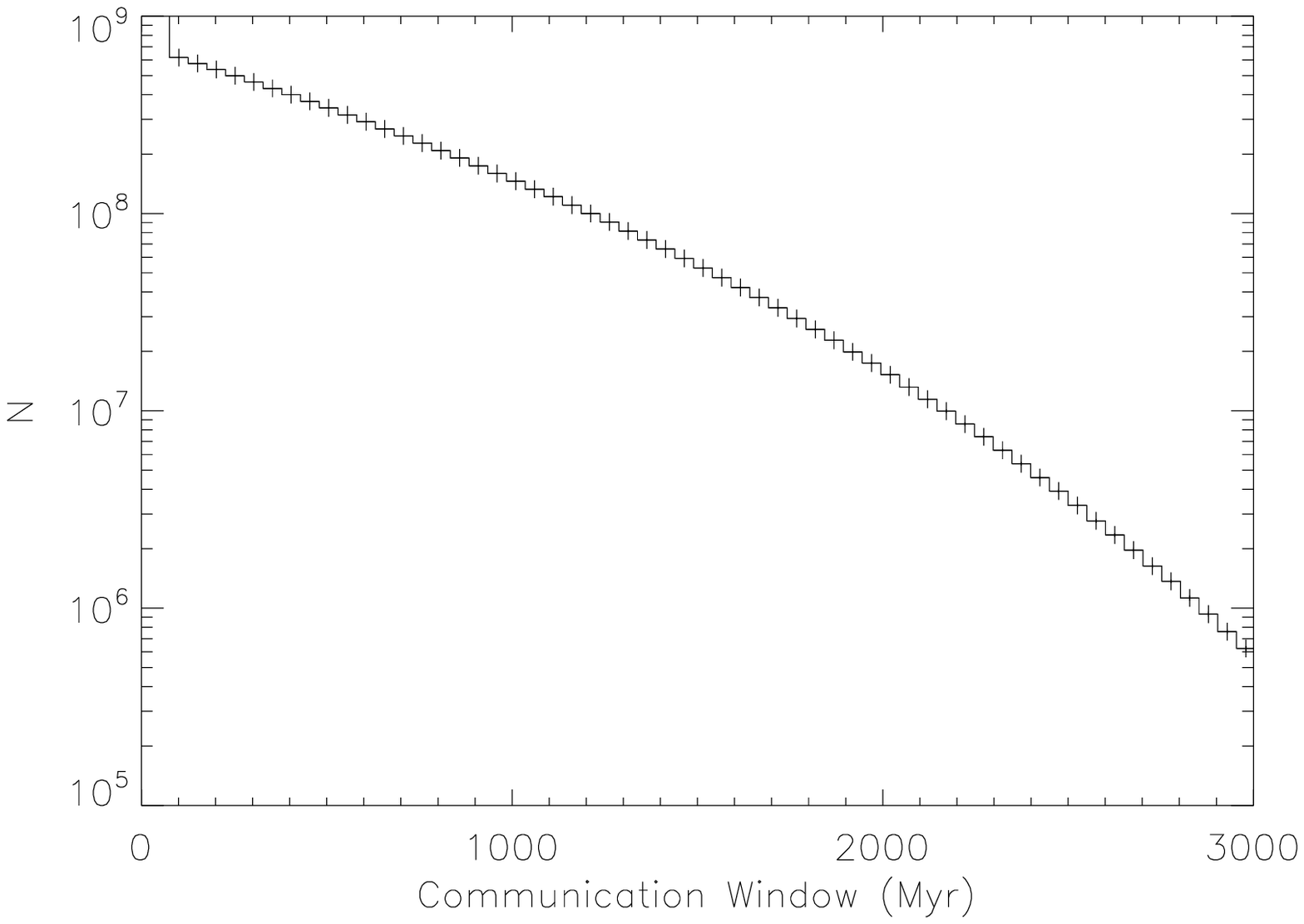} &
\includegraphics[scale = 0.5]{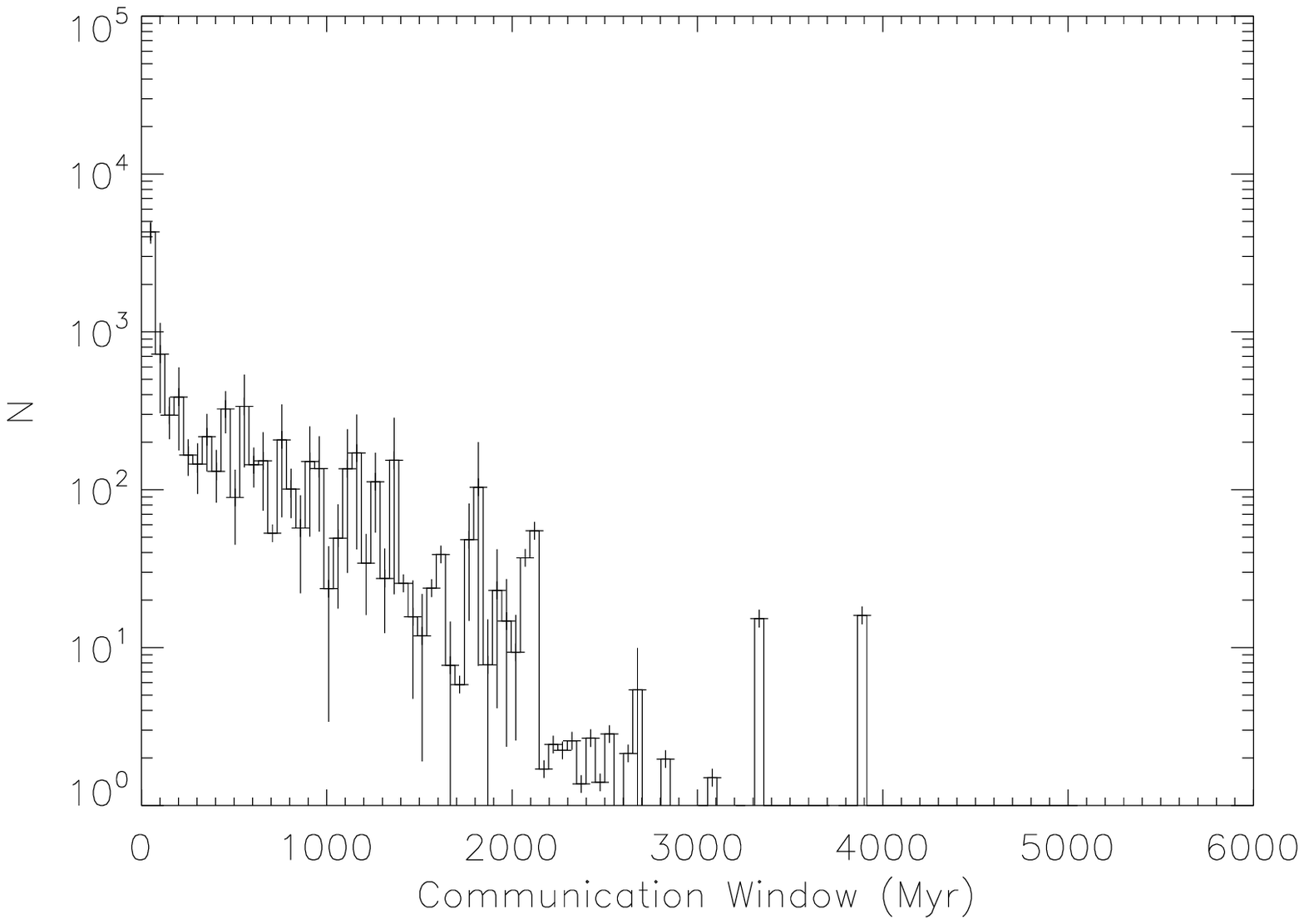} \\
\end{array}$
\caption{\emph{Communications Window (maximum time interval for communication) for ICPs for the Baseline Hypothesis (left) and the Rare Earth Hypothesis (right).} \label{fig:dt}}
\end{center}
\end{figure*}

\begin{figure*}
\begin{center}$
\begin{array}{cc}
\includegraphics[scale = 0.5]{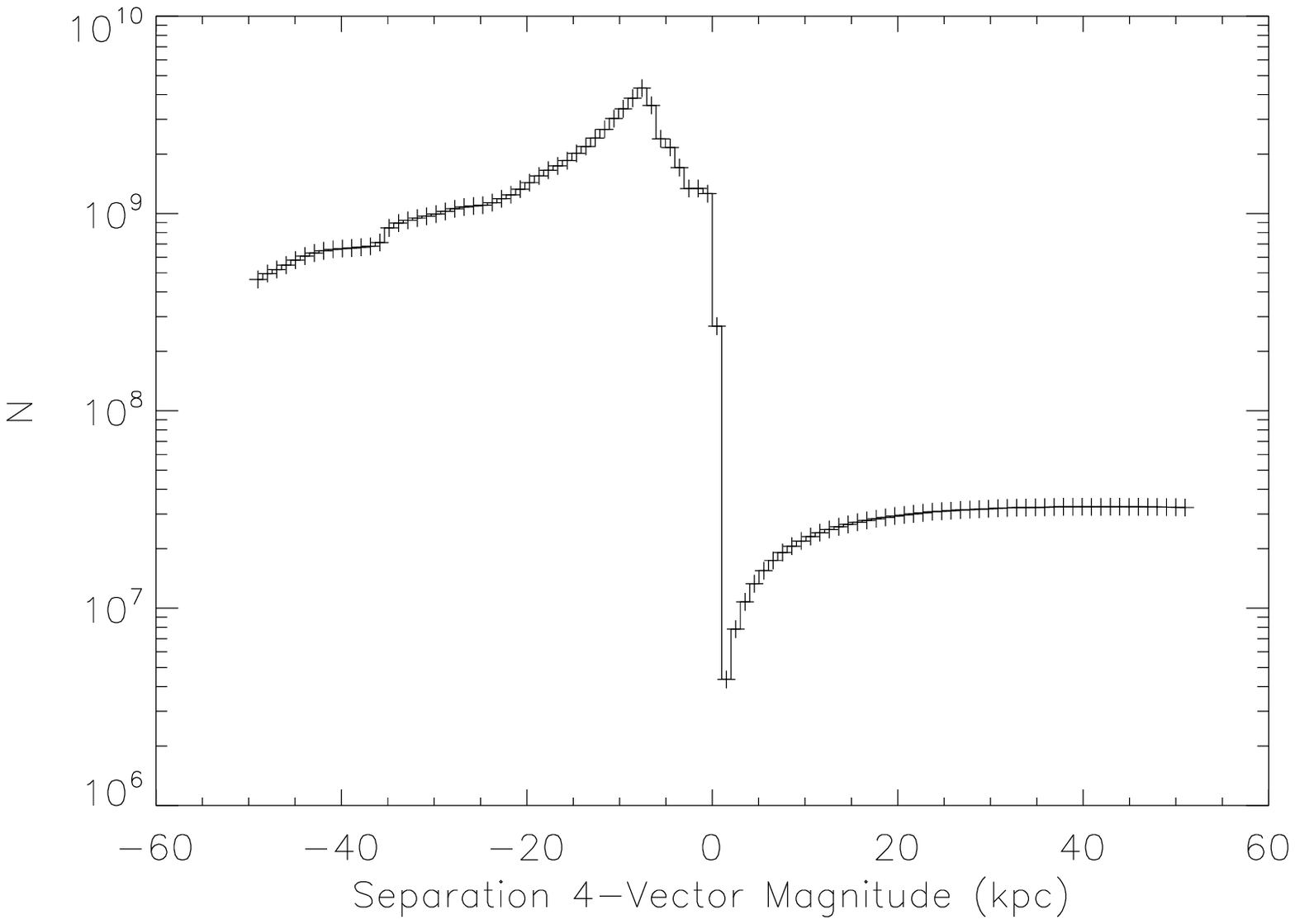} &
\includegraphics[scale = 0.5]{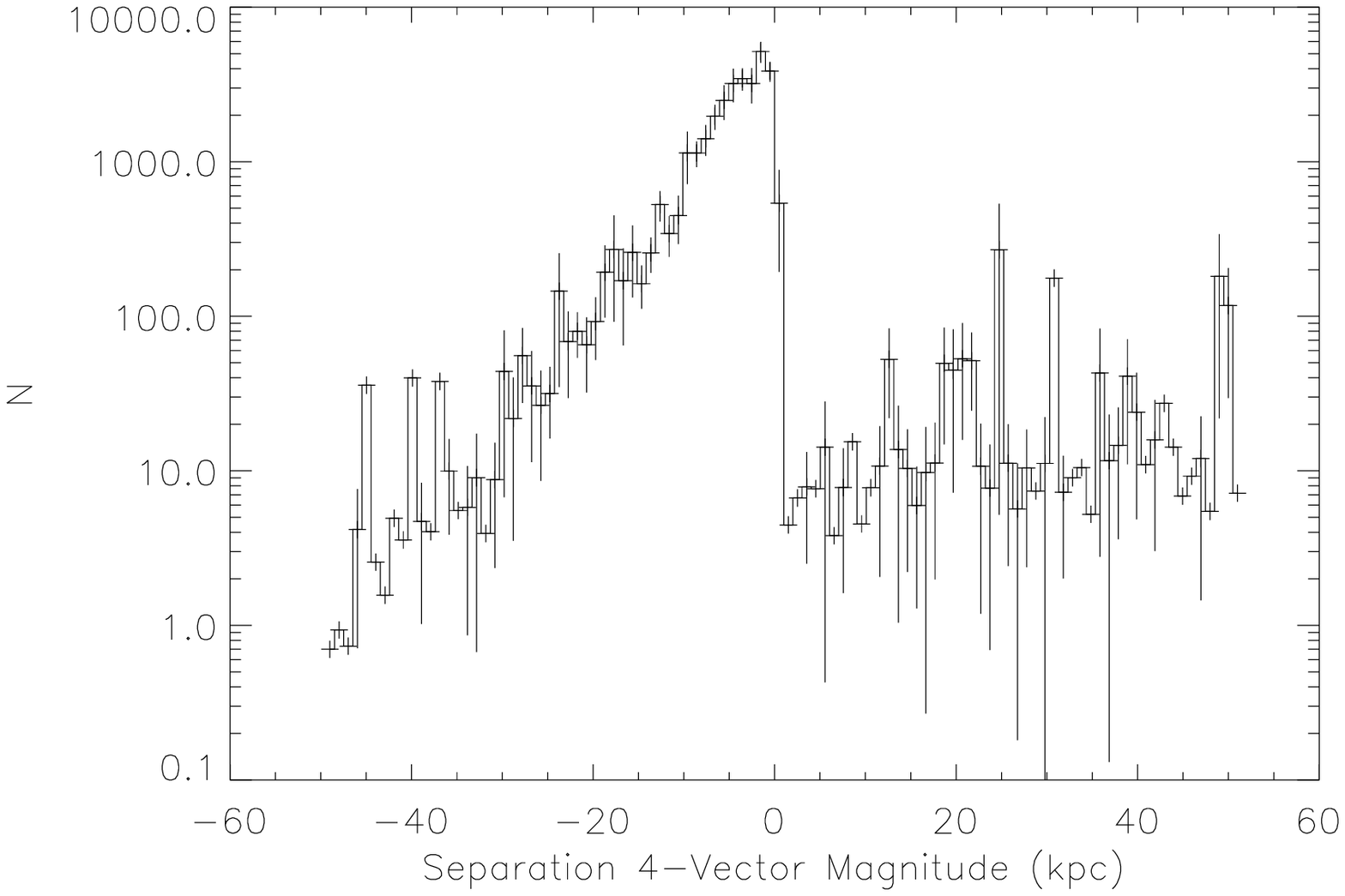} \\
\end{array}$
\caption{\emph{Space-time interval ($ds^2$) of ICPs for the Baseline Hypothesis (left) and the Rare Earth Hypothesis (right).  Unconnected ICPs have $ds^2 <0$, connected civilisations have $ds^2 \geq 0$. } \label{fig:d4x}}
\end{center}
\end{figure*}

\begin{figure*}
\begin{center}$
\begin{array}{cc}
\includegraphics[scale = 0.5]{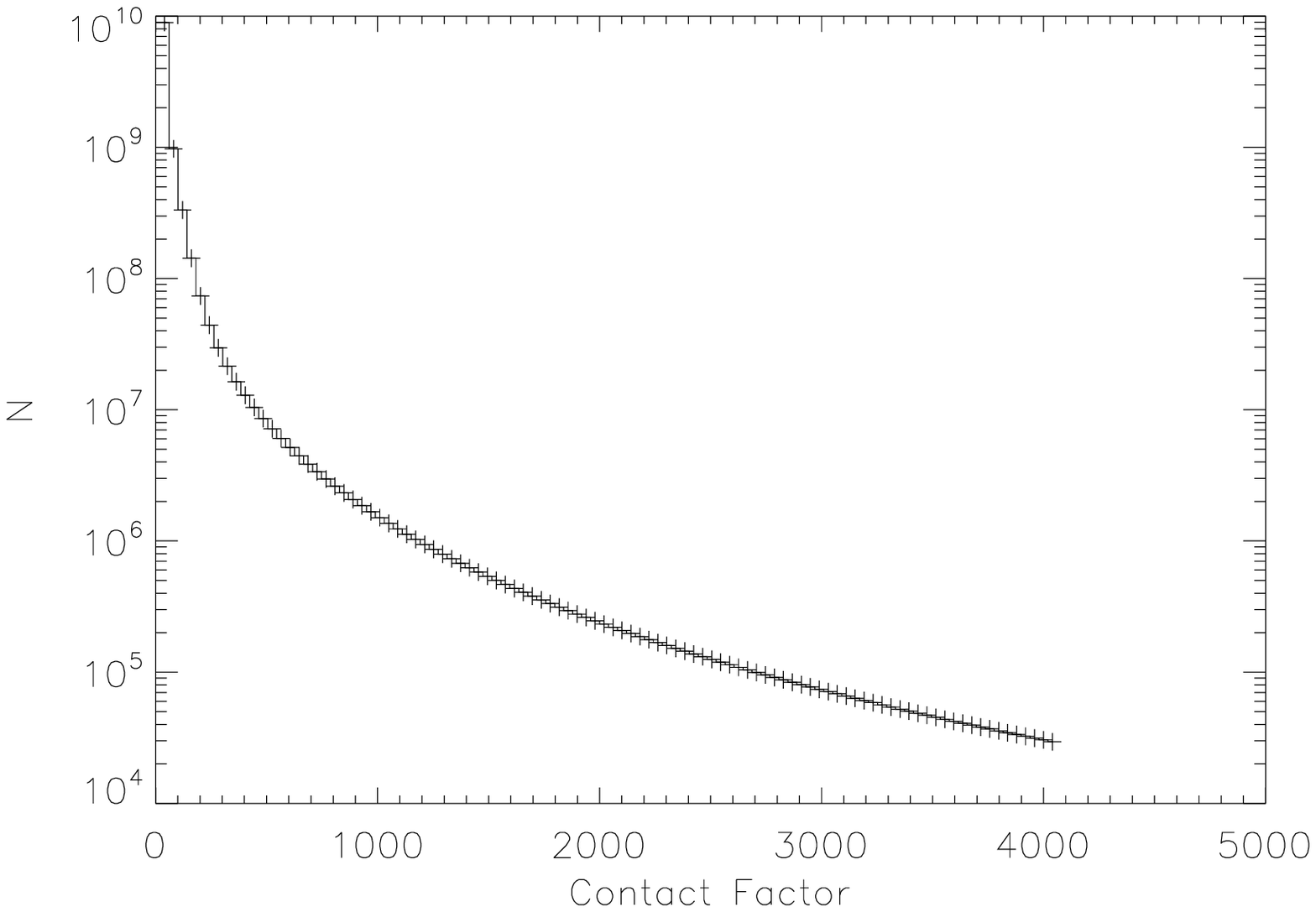} &
\includegraphics[scale = 0.5]{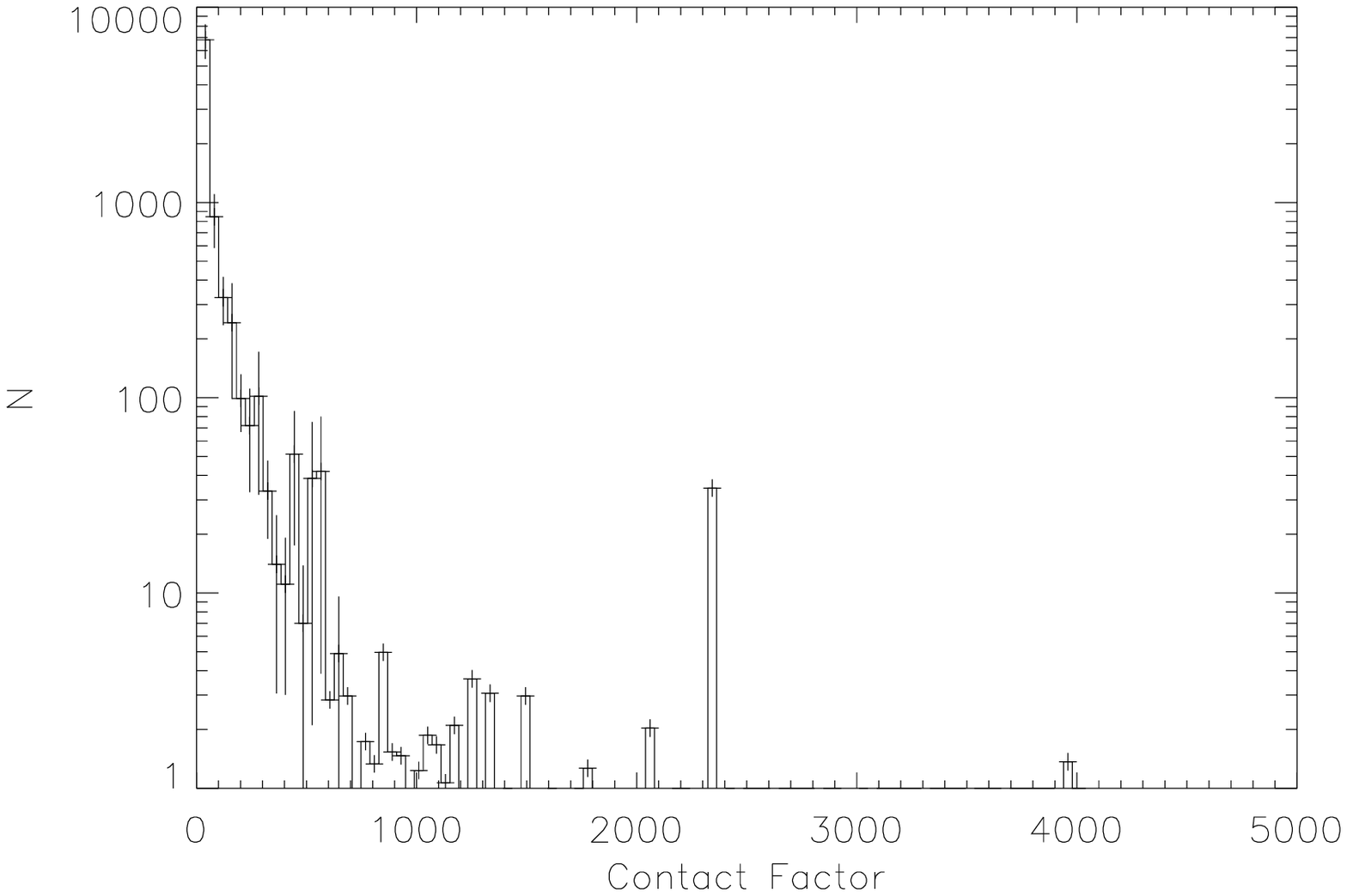} \\
\end{array}$
\caption{\emph{Contact Factor (number of exchanged signal pairs) for ICPs for the Baseline Hypothesis (left) and the Rare Earth Hypothesis (right).} \label{fig:c_fac}}
\end{center}
\end{figure*}

\noindent Does this reduced separation imply increased connectivity? The answer depends on the communication window for the ICP.  The longer the window (i.e. the larger overlap in history where both civilisations exist), the longer the separation can be while allowing the ICP to be connected.  The Baseline Hypothesis favours shorter communication windows (Figure \ref{fig:dt}), which reduces the connectivity.  Apart from small fluctuations at larger values, the Rare Earth Hypothesis agrees.  When considering the space-time interval $ds^2$ (Figure \ref{fig:d4x}), the reduced connectivity becomes apparent.  ICPs that are unconnected (negative values) are much more frequent than connected civilisations (positive or zero values).  This does not spell the end for SETI, however, when the contact factor (i.e. the number of conversations) is considered: although few ICPs enjoy the privilege of contacting each other, those that do can expect a great deal of conversation (Figure \ref{fig:c_fac}), where each hypothesis agrees that a select few will enjoy potentially thousands of exchanges with other civilisations.

\subsection{Future Improvements to the Model}

\noindent Numerical modelling of this type is generally a shadow of the entity it attempts to model, in this case the Milky Way and its constituent stars, planets and other objects.  While a substantial improvement over the work of Paper I, there is still potential for future work.  Several suggestions are listed here.

\begin{enumerate}
\item A more accurate Galactic model, taking into better account its chemical diversity, stellar clustering and the inner regions (specifically the central supermassive black hole and the hypervelocity stars orbiting it).
\item An improved planetary architecture model, better equipped to deal with moons and the planet mass-semimajor axis distribution.  Also missing is the modelling of orbital eccentricity and inclination, potentially of great importance in issues of habitability (e.g. \citealt{Williams_and_Pollard_02,Spiegel_et_al_08}).
\item Improved modelling of the connectivity of civilisations (potentially extending to the modelling of interstellar colonisation and face-to-face contact).
\end{enumerate}

\section{Conclusions}\label{sec:Conclusions}

\noindent This paper has tested the Rare Earth Hypothesis \citep{rare_Earth} using the Monte Carlo Realisation techniques outlined in \citet{mcseti1}.  By comparing the results to a baseline hypothesis, the influences of the criteria for a planet to be officially designated as ``an Earthlike planet" can be studied.  In this work, the criteria were limited to planet mass, star mass, the presence of a moon, and the presence of a Jupiter type object in a more distant orbit.  It is shown that these criteria alone greatly reduce the number of intelligent civilisations in the Galaxy (compared to the baseline).  As expected, the stellar mass criterion results in a narrow range of planet semi-major axes where intelligent biospheres exist.  Interestingly, the Galactic Habitable Zone, apparent in the Baseline Hypothesis, was not visible in the Rare Earth Hypothesis.   \\

\noindent This result is important for civilisation connectivity: reducing the civilisation separation means that, for a given time interval of communication, civilisations under the Rare Earth Hypothesis are able to exchange more signals than civilisations under more contact-optimistic hypotheses.  The implications for SETI are somewhat mixed: while Earth may be much more likely to be a disconnected than connected civilisation, if it is connected, it can expect substantial conversation from other civilisations (while the Sun remains in the Main Sequence).  Therefore, the Rare Earth Hypothesis (in the formulation described in this work) is a ``soft'' or ``exclusive'' hypothesis (using the nomenclature of \citet{Brin_83,fermi_review}), in that it is not a completely contact-pessimistic hypothesis, but one that is contact optimistic for a small subset of civilisations in the Galaxy.  

\section{Acknowledgements}

\noindent This work has made use of the resources provided by the Edinburgh Compute and Data Facility (ECDF, http://www.ecdf.ed.ac.uk/). The ECDF is partially supported by the eDIKT initiative (http://www.edikt.org.uk).

\bibliographystyle{mn2e} % (must include a bibliography style)
\bibliography{IJAduncanforgan}

\begin{thebibliography}{}

\bibitem[\protect\citeauthoryear{{Annis}}{{Annis}}{1999}]{Annis}
{Annis} J.,  1999, Journal of the British Interplanetary Society, 52, 19

\bibitem[\protect\citeauthoryear{{Armitage}}{{Armitage}}{2007}]{Armitage_mig}
{Armitage} P.~J.,  2007, \apj, 665, 1381

\bibitem[\protect\citeauthoryear{{Batygin} \& {Laughlin}}{{Batygin} \&
  {Laughlin}}{2008}]{Batgyin_and_Laughlin_08}
{Batygin} K.,  {Laughlin} G.,  2008, \apj, 683, 1207

\bibitem[\protect\citeauthoryear{{Bounamam Christine}, {von Bloh} \&
  {Franck}}{{Bounamam Christine} et~al.}{2007}]{how_rare}
{Bounamam Christine} {von Bloh} W.,    {Franck} S.,  2007, Astrobiology, 7, 745

\bibitem[\protect\citeauthoryear{{Brin}}{{Brin}}{1983}]{Brin_83}
{Brin} G.~D.,  1983, \qjras, 24, 283

\bibitem[\protect\citeauthoryear{{Canfield}}{{Canfield}}{2005}]{earth_oxygen}
{Canfield} D.~E.,  2005, Annual Review of Earth and Planetary Sciences, 33, 1

\bibitem[\protect\citeauthoryear{{Carr}, {Belton}, {Chapman}, {Davies},
  {Geissler}, {Greenberg}, {McEwen}, {Tufts}, {Greeley}, {Sullivan}
  et~al.,}{{Carr} et~al.}{1998}]{ocean_europa}
{Carr} M.~H.,  {Belton} M.~J.~S.,  {Chapman} C.~R.,  {Davies} M.~E.,
  {Geissler} P.,  {Greenberg} R.,  {McEwen} A.~S.,  {Tufts} B.~R.,  {Greeley}
  R.,  {Sullivan} R.,    et~al., 1998, \nat, 391, 363

\bibitem[\protect\citeauthoryear{{Carter}}{{Carter}}{2008}]{stages}
{Carter} B.,  2008, International Journal of Astrobiology, 7, 177

\bibitem[\protect\citeauthoryear{{Cavicchioli}}{{Cavicchioli}}{2002}]{extremop%
hiles}
{Cavicchioli} R.,  2002, Astrobiology, 2, 281

\bibitem[\protect\citeauthoryear{{Cirkovic}}{{Cirkovic}}{2009}]{fermi_review}
{Cirkovic} M.~M.,  2009, Serbian Astronomical Journal, 178, 1

\bibitem[\protect\citeauthoryear{{Cresswell} \& {Nelson}}{{Cresswell} \&
  {Nelson}}{2006}]{resonances}
{Cresswell} P.,  {Nelson} R.~P.,  2006, \aap, 450, 833

\bibitem[\protect\citeauthoryear{{Diaz} \& {Schulze-Makuch}}{{Diaz} \&
  {Schulze-Makuch}}{2006}]{extremophiles_2}
{Diaz} B.,  {Schulze-Makuch} D.,  2006, Astrobiology, 6, 332

\bibitem[\protect\citeauthoryear{{Ford} \& {Rasio}}{{Ford} \&
  {Rasio}}{2008}]{Ford_and_Rasio_08}
{Ford} E.~B.,  {Rasio} F.~A.,  2008, \apj, 686, 621

\bibitem[\protect\citeauthoryear{Forgan}{Forgan}{2009}]{mcseti1}
Forgan D.,  2009, International Journal of Astrobiology, 8, 121

\bibitem[\protect\citeauthoryear{Formisano, Atreya, Encrenaz, Ignatiev \&
  Giuranna}{Formisano et~al.}{2004}]{methane_mars}
Formisano V.,  Atreya S.,  Encrenaz T.,  Ignatiev N.,    Giuranna M.,  2004,
  Science, 306, 1758

\bibitem[\protect\citeauthoryear{{Hart}}{{Hart}}{1979}]{Hart_HZ}
{Hart} M.~H.,  1979, Icarus, 37, 351

\bibitem[\protect\citeauthoryear{{Horner} \& {Jones}}{{Horner} \&
  {Jones}}{2008}]{jupiter_foe_1}
{Horner} J.,  {Jones} B.~W.,  2008, International Journal of Astrobiology, 7,
  251

\bibitem[\protect\citeauthoryear{{Horner} \& {Jones}}{{Horner} \&
  {Jones}}{2009}]{jupiter_foe_2}
{Horner} J.,  {Jones} B.~W.,  2009, International Journal of Astrobiology, 8,
  75

\bibitem[\protect\citeauthoryear{{Horner}, {Jones} \& {Chambers}}{{Horner}
  et~al.}{2010}]{jupiter_foe_3}
{Horner} J.,  {Jones} B.~W.,    {Chambers} J.,  2010, International Journal of
  Astrobiology, 9, 1

\bibitem[\protect\citeauthoryear{{Ida} \& {Lin}}{{Ida} \&
  {Lin}}{2008}]{Ida_Lin_mig}
{Ida} S.,  {Lin} D.~N.~C.,  2008, \apj, 673, 487

\bibitem[\protect\citeauthoryear{{Kasting}, {Whitmire} \& {Reynolds}}{{Kasting}
  et~al.}{1993}]{Kasting_et_al_93}
{Kasting} J.~F.,  {Whitmire} D.~P.,    {Reynolds} R.~T.,  1993, Icarus, 101,
  108

\bibitem[\protect\citeauthoryear{{Kipping}, {Fossey} \& {Campanella}}{{Kipping}
  et~al.}{2009}]{Kipping_moon}
{Kipping} D.~M.,  {Fossey} S.~J.,    {Campanella} G.,  2009, \mnras, pp 1364--+

\bibitem[\protect\citeauthoryear{{Krasnopolsky}, {Maillard} \&
  {Owen}}{{Krasnopolsky} et~al.}{2004}]{methane_mars_2}
{Krasnopolsky} V.~A.,  {Maillard} J.~P.,    {Owen} T.~C.,  2004, Icarus, 172,
  537

\bibitem[\protect\citeauthoryear{{Lineweaver}, {Fenner} \&
  {Gibson}}{{Lineweaver} et~al.}{2004}]{GHZ}
{Lineweaver} C.~H.,  {Fenner} Y.,    {Gibson} B.~K.,  2004, Science, 303, 59

\bibitem[\protect\citeauthoryear{{Maccone}}{{Maccone}}{2009}]{stat_drake}
{Maccone} C.,  2009, Acta Astronautica, in press

\bibitem[\protect\citeauthoryear{{Ostlie} \& {Carroll}}{{Ostlie} \&
  {Carroll}}{1996}]{Ostlie_and_Caroll}
{Ostlie} D.~A.,  {Carroll} B.~W.,  1996, {An Introduction to Modern Stellar
  Astrophysics}

\bibitem[\protect\citeauthoryear{{Paardekooper} \& {Papaloizou}}{{Paardekooper}
  \& {Papaloizou}}{2008}]{Paardekooper_and_Papaloizou_08}
{Paardekooper} S.-J.,  {Papaloizou} J.~C.~B.,  2008, \aap, 485, 877

\bibitem[\protect\citeauthoryear{{Parkinson}, {Liang}, {Hartman}, {Hansen},
  {Tinetti}, {Meadows}, {Kirschvink} \& {Yung}}{{Parkinson}
  et~al.}{2007}]{plume_enceladus}
{Parkinson} C.~D.,  {Liang} M.-C.,  {Hartman} H.,  {Hansen} C.~J.,  {Tinetti}
  G.,  {Meadows} V.,  {Kirschvink} J.~L.,    {Yung} Y.~L.,  2007, \aap, 463,
  353

\bibitem[\protect\citeauthoryear{{Prialnik}}{{Prialnik}}{2000}]{Prialnik}
{Prialnik} D.,  2000, {An Introduction to the Theory of Stellar Structure and
  Evolution}

\bibitem[\protect\citeauthoryear{{Raup} \& {Sepkoski}}{{Raup} \&
  {Sepkoski}}{1982}]{Raup_and_Sepkoski}
{Raup} D.~M.,  {Sepkoski} J.~J.,  1982, Science, 215, 1501

\bibitem[\protect\citeauthoryear{{Raymond}, {Armitage} \& {Gorelick}}{{Raymond}
  et~al.}{2009}]{planet_planet}
{Raymond} S.~N.,  {Armitage} P.~J.,    {Gorelick} N.,  2009, \apjl, 699, L88

\bibitem[\protect\citeauthoryear{{Raymond}, {Mandell} \&
  {Sigurdsson}}{{Raymond} et~al.}{2006}]{Earth_after_giants}
{Raymond} S.~N.,  {Mandell} A.~M.,    {Sigurdsson} S.,  2006, Science, 313,
  1413

\bibitem[\protect\citeauthoryear{{Rocha-Pinto}, {Maciel}, {Scalo} \&
  {Flynn}}{{Rocha-Pinto} et~al.}{2000}]{Rocha_Pinto_AMR}
{Rocha-Pinto} H.~J.,  {Maciel} W.~J.,  {Scalo} J.,    {Flynn} C.,  2000, \aap,
  358, 850

\bibitem[\protect\citeauthoryear{{Rocha-Pinto}, {Scalo}, {Maciel} \&
  {Flynn}}{{Rocha-Pinto} et~al.}{2000}]{Rocha_Pinto_SFH}
{Rocha-Pinto} H.~J.,  {Scalo} J.,  {Maciel} W.~J.,    {Flynn} C.,  2000, \aap,
  358, 869

\bibitem[\protect\citeauthoryear{{Rolleston}, {Smartt}, {Dufton} \&
  {Ryans}}{{Rolleston} et~al.}{2000}]{z_grad}
{Rolleston} W.~R.~J.,  {Smartt} S.~J.,  {Dufton} P.~L.,    {Ryans} R.~S.~I.,
  2000, \aap, 363, 537

\bibitem[\protect\citeauthoryear{{Sartoretti} \& {Schneider}}{{Sartoretti} \&
  {Schneider}}{1999}]{moon_detect}
{Sartoretti} P.,  {Schneider} J.,  1999, \aaps, 134, 553

\bibitem[\protect\citeauthoryear{{Schr{\"o}der} \& {Connon
  Smith}}{{Schr{\"o}der} \& {Connon Smith}}{2008}]{Schroder_and_Smith_08}
{Schr{\"o}der} K.-P.,  {Connon Smith} R.,  2008, \mnras, 386, 155

\bibitem[\protect\citeauthoryear{{Spencer} \& {Grinspoon}}{{Spencer} \&
  {Grinspoon}}{2007}]{plume_enceladus_2}
{Spencer} J.,  {Grinspoon} D.,  2007, \nat, 445, 376

\bibitem[\protect\citeauthoryear{{Spiegel}, {Menou} \& {Scharf}}{{Spiegel}
  et~al.}{2008}]{Spiegel_et_al_08}
{Spiegel} D.~S.,  {Menou} K.,    {Scharf} C.~A.,  2008, \apj, 681, 1609

\bibitem[\protect\citeauthoryear{{Stofan}, {Elachi}, {Lunine}, {Lorenz},
  {Stiles}, {Mitchell}, {Ostro}, {Soderblom}, {Wood}, {Zebker}
  et~al.,}{{Stofan} et~al.}{2007}]{lake_titan}
{Stofan} E.~R.,  {Elachi} C.,  {Lunine} J.~I.,  {Lorenz} R.~D.,  {Stiles} B.,
  {Mitchell} K.~L.,  {Ostro} S.,  {Soderblom} L.,  {Wood} C.,  {Zebker} H.,
  et~al., 2007, \nat, 445, 61

\bibitem[\protect\citeauthoryear{{Vukotic} \& {Cirkovic}}{{Vukotic} \&
  {Cirkovic}}{2007}]{Vukotic_and_Cirkovic_07}
{Vukotic} B.,  {Cirkovic} M.~M.,  2007, Serbian Astronomical Journal, 175, 45

\bibitem[\protect\citeauthoryear{{Vukotic} \& {Cirkovic}}{{Vukotic} \&
  {Cirkovic}}{2008}]{Vukotic_and_Cirkovic_08}
{Vukotic} B.,  {Cirkovic} M.~M.,  2008, Serbian Astronomical Journal, 176, 71

\bibitem[\protect\citeauthoryear{{Waltham}}{{Waltham}}{2004}]{anthropic_moon}
{Waltham} D.,  2004, Astrobiology, 4, 460

\bibitem[\protect\citeauthoryear{{Ward} \& {Brownlee}}{{Ward} \&
  {Brownlee}}{2000}]{rare_Earth}
{Ward} P.,  {Brownlee} D.,  2000, {Rare Earth : Why Complex Life is Uncommon in
  the Universe}

\bibitem[\protect\citeauthoryear{{Williams} \& {Pollard}}{{Williams} \&
  {Pollard}}{2002}]{Williams_and_Pollard_02}
{Williams} D.~M.,  {Pollard} D.,  2002, International Journal of Astrobiology,
  1, 61

\bibitem[\protect\citeauthoryear{{Wyatt}, {Clarke} \& {Greaves}}{{Wyatt}
  et~al.}{2007}]{Wyatt_Z}
{Wyatt} M.~C.,  {Clarke} C.~J.,    {Greaves} J.~S.,  2007, \mnras, 380, 1737

\end{thebibliography}

\end{document}